\documentclass[pra,preprint]{revtex4-1}
\usepackage{amsmath,amssymb}
\usepackage{bbm}
\usepackage{mathrsfs}
\usepackage{graphicx}
\usepackage{xcolor}
\newcommand{\etal}{\textit{et al.}}

\begin{document}
\title{Emergence of prethermal states in a driven dissipative system through cross-correlated dissipation} 

\author{Arnab Chakrabarti} 
\email{arnab.chakrabarti@weizmann.ac.il}
\affiliation
{AMOS and Department of Chemical and Biological Physics, Weizmann Institute of Science,
Rehovot - 7610001,
Israel }

\author{Rangeet Bhattacharyya}
\email{rangeet@iiserkol.ac.in}
\affiliation
{Department of Physical Sciences,
Indian Institute of Science~Education and Research Kolkata, 
Mohanpur - 741246, West Bengal, India }
 
\date{\today}

\begin{abstract}
Periodically driven closed quantum many-body systems are known to exhibit prethermal or
quasi-steady-state dynamics. In this work, we theoretically show that such prethermal phases can appear in
the dynamics of a dipolar two-spin-$1/2$ system coupled to a heat bath if the cross terms between the drive
and dipolar interactions are taken into consideration. To this end, we use our recently-reported
fluctuation-regulated quantum master equation [A. Chakrabarti and R. Bhattacharyya, Phys. Rev. A 97, 063837
(2018)], to show that the predicted dynamics can successfully explain the experimentally observed features
of the transient and prethermal regime.
\end{abstract}

\maketitle

\section{Introduction}

Periodically driven quantum systems appear in a large class of problems of interest and, therefore, are the
subject of continued investigation \cite{eckardt17,fleckenstein21, beatrez21, dalessio14, sen21}.
Theoretical and experimental endeavors have proved that there exist three different regimes in the dynamics
of periodically driven quantum ensembles, v.i.z. i) a transient phase, ii) a prethermal quasi-steady state,
and iii) unconstrained thermalization when the drive amplitude or the Rabi frequency is sufficiently high
\cite{beatrez21, santos21}. Similar dynamical features in the presence of drives having lower amplitudes
have also been predicted \cite{fleckenstein21}. The emergence of the prethermal quasi-steady state is of
particular importance, which can be used for engineering quantum gates, preserving coherences (and hence
quantum information), and understanding the physics of thermalization processes in general
\cite{goldman14,bukov15,singh19}. Most of the theoretical framework developed in this regard concerns closed
quantum ensembles whereby unconstrained thermalization results from the breakdown of a Floquet-Magnus
approximation used to describe the transient and prethermal dynamics \cite{dalessio14}. Only recently, the
emergence of a prethermal quasi-steady state has been predicted by Angl{\'{e}}s-Castillo et al. for a
two-level system, coupled in cascade to two distinct thermal reservoirs, with different equilibrium
temperatures \cite{castillo20}. The quasi-stable prethermal state observed therein is due to local
thermalization induced by the reservoir to which the system is directly coupled, while the final
nonequilibrium steady state corresponds to global equilibration \cite{castillo20}. It is then pertinent to
ask whether prethermal states can also be observed in multi-partite open quantum systems that are directly
coupled to a thermal bath?

In this context, it is interesting to note that a quantum system comprising interacting sub-parts,
coupled to a single external bath (source of decoherence), can have additional immunity to decoherence
\cite{grigorenko05}. Moreover, it has already been demonstrated that the cooperative dynamics of two
interacting (dipole-dipole) two-level atoms can result in the inhibition of their fluorescence -- a
phenomenon attributed to the coupling between the symmetric and antisymmetric collective states
\cite{lawande90}. The emergence of collective steady-states of driven two-level systems coupled by their
mutual dipolar interactions has received considerable interest in the recent years \cite{ parmee17,
parmee18, parmee20, landa20}. In this work, our aim is to show that the prethermal phase can also appear in
the dynamics of an interacting two-qubit system, coupled to a thermal bath if we consider the
cross-correlated dissipators from the drive and inter-qubit interactions. Such prethermal states are essentially
quasi-stable collective coherences, which emerge due to their relative immunity to some of the decay
channels.
     
The standard techniques for treating open quantum dynamics cannot be adopted to account for the interplay
of drive and inter-qubit interaction, which we wish to capture. But, our recently-proposed Fluctuation
Regulated Quantum Master Equation (FRQME), which derives all second-order terms with an explicit regulator
originating from the average effect of thermal fluctuations in the environment, can offer a probable
solution \cite{chakrabarti2018b}. Through the second-order terms of a coherent drive, FRQME predicts the
presence of a unique drive-dependent decay rate, which has been experimentally observed by the authors in a
single spin ensemble \cite{chakrabarti2018b, chakrabarti2018a}. Motivated by this success, in the present
work, we use FRQME to describe the dissipative dynamics of a driven two-spin-$1/2$ system having 
dipolar interactions. Interestingly, the drive-dependence of thermalization rates has recently been
reported for periodically driven closed quantum many-body systems which show prethermal phases in their
dynamics \cite{fleckenstein21}. 

To focus mainly on the effect of drive-dipole cross-terms in the second order, we shall assume a weak
system-environment coupling -- weaker than both the drive and dipolar interaction. The equations of motion
obtained from this exercise suggest that an initially created collective coherence will be persistent. In
practice, the predicted persistent coherence will decay due to the system-environment coupling, indicating
the transition to complete thermalization. The quasi-steady-state coherences are akin to the spin-locked
magnetization often encountered in magnetic resonance.

The manuscript is organized in the following order: in the next section, we briefly describe the FRQME,
pointing out its key features. We then apply this FRQME to a driven two-spin-$1/2$ system, coupled by
their mutual dipolar interactions, whereby we obtain drive-dipole cross-terms in
the dissipator. In the following section, we present the relevant macroscopic dynamical equations to
describe the observed phenomena. The solution of these dynamical equations illustrates the emergence of the
collective steady-state coherence and can account for previously observed experimental results, unlike other
theoretical approaches. We end with a discussion on the method and the results obtained and a short conclusion
highlighting the implications and probable applications of this approach.

\section{Fluctuation-regulated quantum master equation}

One of the major motivations of our alternate formulation of the quantum master equation was to include the
higher-order effects of external drives in the dynamics. Such higher-order effects influence the dynamics
through well-studied shift terms (such as, light shifts and Bloch-Siegert shifts), and relatively less
explored drive-induced dissipation terms \cite{chatterjee2020, chakrabarti2018a, chanda2020, chanda2021}.
Since the complete derivation and the essential features of FRQME have been presented elsewhere
\cite{chakrabarti2018b}, here we present a brief review of its framework. The basic premises of a standard
Markovian QME including the Born and Markov approximations, are used in our formulation of FRQME. In
addition to those, we also take into account an additional process in the form of an explicit Hamiltonian
which strives to capture the ubiquitous thermal fluctuations in the bath. \textcolor{black}{Specifically, we
formulate our problem for a quantum system (ensemble) that is interacting with a thermal bath. Each member
of the system ensemble (each $2$-spin unit in our case) is directly coupled to a finite portion of the bath,
which we name as the ``local environment". For example, in a dilute spin-ensemble (as in \cite{beatrez21}),
each spin unit ($2$-spin unit in our case) is directly coupled to the spatial degrees of freedom of
molecules in its immediate vicinity. The collection of all these local environments form the bath, which is
in thermal equilibrium. The time-independent equilibrium density matrix of the bath is obtained from an
ensemble average of the local-environment density matrices. Individual local environments must always
experience equilibrium thermal fluctuations in order to ensure that there is no average coherence build-up
in the bath, even though evolution under system-local environment interaction takes place.}

So, for a system weakly-coupled to its local environment, the full Hamiltonian \textcolor{black}{of each
ensemble member (system + local environment)}, in units of angular frequency, is given by

\begin{equation}\label{hamgen}
   \mathscr{H}(t) = \mathscr{H}_S^{\circ} + \mathscr{H}_L^{\circ} + \mathscr{H}_{SL} + \mathscr{H}_{\rm sys}(t) + \mathscr{H}_L(t),
\end{equation}
where, $\mathscr{H}_S^{\circ}$ and $\mathscr{H}_L^{\circ}$ denote the bare Hamiltonians of the system and
the local environment, respectively. $\mathscr{H}_{SL}$ is the coupling between the system and its local
environment while $\mathscr{H}_{\rm sys}(t)$ includes all other terms affecting the system alone (e.g.
drive, inter-spin interactions in spin networks etc.). Thus, $\mathscr{H}_{\rm sys}(t)$ is assumed to be
time-dependent in general. The explicitly time-dependent term $\mathscr{H}_L(t)$ represents the fluctuations
in the local environments. The collection of these local environments constitute the heat bath, which is
assumed to be in thermal equilibrium at an inverse temperature $\beta$, while its energy levels are defined
by $\mathscr{H}_L^{\circ}$. Since, thermal fluctuations should not destroy the equilibrium populations of
the energy levels, $\mathscr{H}_L(t)$ is chosen to be diagonal in the eigen-basis $\lbrace \vert
\phi_j\rangle \rbrace $ of $\mathscr{H}_L^{\circ}$: $\mathscr{H}_L(t) = \sum\limits_j f_j(t)\vert
\phi_j\rangle\langle \phi_j \vert$, where, $f_j(t)$-s are modeled as independent, Gaussian,
$\delta$-correlated stochastic variables with zero mean and standard deviation $\kappa$ i.e.
$\overline{f_j(t)} = 0$ and $\overline{f_j(t_1)f_k(t_2)} = \kappa^2\delta_{jk}\delta(t_1 - t_2)$ (the
overhead line denotes ensemble averaging). \textcolor{black}{Thus, $\mathscr{H}_L(t)$ describes equilibrium
fluctuations in individual local environments and it has been constructed in such a way that the thermal
density matrix remains time-independent as required for sustained equilibrium.}

The derivation of FRQME relies on a time coarse-graining method in the interaction representation of
$\mathscr{H}_S^{\circ} + \mathscr{H}_L^{\circ}$, adequately outlined by Cohen-Tannoudji \etal
\cite{cotandurogryn04,chakrabarti2018b}, in order to smoothen out the instantaneous effects of the
fluctuations while retaining its average effect in the dynamics. Defining \textcolor{black}{$H_{\rm eff}(t)
:= H_{\rm sys}(t) + H_{SL}(t) $}, where the symbol $H$ with relevant subscripts denote the corresponding
Hamiltonians in the interaction representation, and $\rho_S(t)$ as the reduced density matrix of the system
under study, the FRQME is given by \cite{chakrabarti2018b}:
\begin{eqnarray}\label{dd-qme} 
\frac{d}{dt}\rho_{\rm  S}(t)  &=& -i\; {\rm Tr}_{\rm L}\!\Big[H_{\rm 
eff}(t),\, \rho_{\rm  S}(t) \otimes\rho_{\rm  L}^{\rm eq}\Big]^{\rm sec} \nonumber\\
 &\ & -\int_{0}^{\infty}\kern-11pt d\tau\; {\rm Tr}_{\rm L}\!\Big[H_{\rm eff}(t),\,\Big[H_{\rm  eff}(t-\tau), \nonumber\\
 & & \hspace{1.5cm} \rho_{\rm  S}(t)\otimes
\rho_{\rm  L}^{\rm eq}\Big]\Big]^{\rm sec}e^{-{\vert\tau\vert}/{\tau_c}},
\end{eqnarray} 
where, $\rho_{\rm  L}^{\rm eq}$ is the equilibrium density matrix of the bath and we have defined $\tau_c =
2/\kappa^2$, \cite{chakrabarti2018b}. The superscript ``sec" indicates that only secular contributions are
retained \cite{chakrabarti2018b}. \textcolor{black}{ The crucial effect of introducing the local-environment
fluctuations is to obtain this explicit time-scale $\tau_c$ during which all second order terms in the FRQME
remain significant. In standard treatments no explicit time-scale is present, although its presence is
assumed implicitly (see \cite{cotandurogryn04}), only in second-order terms of the interaction. In contrast,
the second-order drive-drive or dipole-dipole or drive-dipole terms in FRQME are regularized by an
exponential memory kernel (with characteristic time $\tau_c$) originating from the finite dephasing of the
local environment, induced by the fluctuations.} Importantly, due to the presence of this regulator in all
second-order terms, FRQME contains second-order contributions of both the spin-environment coupling as well
as the Hamiltonians which act on the system alone e.g. an external drive.  As shown by the authors earlier,
apart from the regular dissipators from the system-bath coupling, this master equation predicts additional
relaxation terms quadratic in the drive amplitude. These drive-induced dissipation terms are Kramers-Kronig
pairs of the familiar light-shift terms \cite{chakrabarti2018b, chakrabarti2018a}. The main advantage of
using FRQME for the two-spin ensemble is that the fluctuation-regulated second-order terms include the cross
terms between the drive and the inter-spin coupling Hamiltonians.

\section{{Persistent two-spin coherence: Prethermal steady state}}

Having introduced the FRQME, we now focus on its application to a two-spin-$1/2$ ensemble, with dipolar
interactions. In this case, $\mathscr{H}^{\circ}_S$ denotes the bare Zeeman Hamiltonian of the two spins,
while 
\textcolor{black}{
\begin{equation}
            H_{\rm sys} = H_{\rm DD} + H_{\rm drive}, 
\end{equation}
}
in the interaction representation \textcolor{black}{of $\mathscr{H}_S^{\circ} + \mathscr{H}_L^{\circ}$}.
Here $H_{\rm DD}$ represents the secular, semi-classical, dipolar Hamiltonian: 
\begin{equation}\label{dd-sec}
            H_{\rm DD} = 
\frac{\omega_d}{4} \left[2\sigma^{(1)}_z\sigma^{(2)}_z - \sigma^{(1)}_x\sigma^{(2)}_x
- \sigma^{(1)}_y\sigma^{(2)}_y\right]
\end{equation}
and $\sigma_{\alpha}\, ,\, \sigma_{\alpha} \,\forall \alpha \in \lbrace x,y,z\rbrace$ denote the Pauli spin
matrices, with superscripts denoting the particle identifiers.
The factor $\omega_d$ represents the strength of the
coupling. We restrict our analysis to the case where the two spins forming the ensemble of interest are
indistinguishable in all respects, having identical Zeeman splittings (Larmor frequencies). A resonant
co-rotating drive is applied to this system, which we represent by the Hamiltonian, 
\textcolor{black}{
\begin{equation}\label{dd-lockham}
H_{\rm drive}(t) = \frac{\omega_1}{2}\left[\sigma^{(1)}_x + \sigma^{(2)}_x\right],
\end{equation}
}
in the interaction representation, where $\omega_1$ denotes the drive amplitude.  \textcolor{black}{The
FRQME (\ref{dd-qme}) for this system can be expressed as}
\begin{eqnarray}\label{dd-2}
\dot{\rho}_S & = & \big[\,\mathcal{L}_1(H_{\rm DD} + H_{\rm drive}) + \mathcal{L}_2(H_{\rm DD} + H_{\rm drive})\nonumber\\ 
             &   &   + \mathcal{L}_2(H_{SL})\,\big]\,\rho_S,
\end{eqnarray}
where, $\mathcal{L}_1$ denotes the first-order Liouvillian [the first term on the r.h.s. of FRQME
(\ref{dd-qme})] and $\mathcal{L}_2$, the second-order Liouvillian [the second term on the r.h.s. of FRQME
(\ref{dd-qme})] with the Hamiltonians in the argument. In the above equation, the overhead dot ``\,$^.$\,"
indicates time-derivative. In deriving (\ref{dd-2}) we have assumed {${\rm Tr}_{\rm L} [\,\rho_{\rm  L}^{\rm
eq} \, H_{SL}\,] = 0$}. It is important to note that $\mathcal{L}_2(H_{\rm DD} + H_{\rm drive})$ includes
auto- as well as cross-terms between $H_{\rm DD}$ and $H_{\rm drive}$.  
\textcolor{black}{$\mathcal{L}_2(H_{SL})$ is the standard decay channel describing Markovian damping of spin
coherences. All previous quantum master equations have this contribution and as such, it cannot predict the
emergence of prethermal quasi-steady states in systems coupled to a single heat bath. The unusual and most
interesting features lie in the decay channel induced by $\mathcal{L}_2(H_{\rm DD} + H_{\rm drive})$, due
the presence of cross-terms between $H_{\rm DD}$ and $H_{\rm drive}$. In order to simplify calculations and
clearly showcase the prethermalization induced by these cross terms, in this work we explore the parameter
regime $\mathcal{L}_2(H_{\rm DD} + H_{\rm drive}) \gg \mathcal{L}_2(H_{SL})$. Note that this does not imply
that we assume a vanishing coupling $H_{\rm SL}$ between the system and the bath. As mentioned in our
original formulation of FRQME, a non-zero system-bath coupling is essential for this construction
\cite{chakrabarti2018b}. The choice of the above parameter regime just means that we are showcasing the
short-time behavior keeping in mind that the well known overall damping is always present. This can always
be done to analyze a particular feature of a complex dynamical system. To mimic realistic experimental data
one can simply introduce an overall exponential damper to the solution obtained with just
$\mathcal{L}_2(H_{\rm DD} + H_{\rm drive})$, or perform the the exact calculation assuming a form of
$\mathcal{L}_2(H_{SL})$. Either case would not change the core physics of  $\mathcal{L}_2(H_{\rm DD} +
H_{\rm drive})$. Moreover, this parameter regime implies that the system-dynamics involves two distinct time
scales originating from the dissipator $\mathcal{L}_2$: a long-time, slow component through $H_{SL}$ and a
short-time, fast component due to $(H_{\rm DD} + H_{\rm drive})$. Thus we expect the two-spin system to
reach a quasi-steady state with respect to the fast term [$\mathcal{L}_2(H_{\rm DD} + H_{\rm drive})$] much
before the effects of the slow term [$\mathcal{L}_2(H_{SL})$] becomes appreciable. This is analogous to
Castillo et al.'s quasi-steady local thermalization followed by a global thermalization of a two-level
system coupled in cascade to two heat baths \cite{castillo20}. Since our aim is to analyze the emergence and
nature of the quasi-steady (prethermal) state of the $2$-spin-$1/2$ ensemble, we restrict our analysis to
the study of $\mathcal{L}_2(H_{\rm DD} + H_{\rm drive})$ in this work.}


\subsection{Dynamical Equations}

The two spin-$1/2$ density matrix ($\rho_S$) is represented by a $4 \times 4$ Hermitian matrix with unit
trace, and hence having $15$ independent matrix elements. While one can solve Eq. (\ref{dd-qme}), for a more
convenient and intuitive description, we recast Eq. (\ref{dd-qme}), in terms of the expectation values of
the observables. One can construct a set of expectation values of a collection of symmetric and
antisymmetric observables using 
\begin{eqnarray}\label{Var}
M^{\pm}_{\alpha} & = & \frac{1}{2}{\rm Tr}\left[\rho_S\left(\sigma^{(1)}_{\alpha} \pm
\sigma^{(2)}_{\alpha}\right)\right],\\
M^{+}_{\alpha\alpha} & = & \frac{1}{4}{\rm Tr}\left[\rho_S\sigma^{(1)}_{\alpha}\sigma^{(2)}_{\alpha}\right],\\
M^{\pm}_{\alpha\beta} & = & \frac{1}{4}{\rm Tr}\left[\rho_S\left(\sigma^{(1)}_{\alpha}\sigma^{(2)}_{\beta}
\pm
\sigma^{(1)}_{\beta}\sigma^{(2)}_{\alpha}\right)\right], \nonumber\\
&  &  \hspace{1cm} \forall \;\; \alpha ,\beta \in \; \lbrace x, y, z \rbrace \; \& \; \alpha \neq \beta.
\end{eqnarray}
where, the superscripts $+$ and $-$ indicate symmetric and antisymmetric combinations of the observables,
respectively.  For the choice of observables, we find that the equations of nine symmetric and six
antisymmetric observables have no cross terms and hence the coefficient matrix is block diagonal. For the
problem at hand, our Hamiltonian is invariant under an exchange of the spin indices. Moreover, we intend to
investigate the dynamics of this system for an initial coherence $M^{+}_{x}$ which is also symmetric with
respect to the exchange of the spin indices. As such, the antisymmetric observables remain zero throughout
the dynamics. Therefore, we show the equations corresponding to only the symmetric observables (we drop the
superscript $+$ from the symmetric observables for clarity).  Using equation (\ref{dd-2}) we arrive at the
following set of differential equations,
\begin{align}\label{dd-9obs}
     & \dot{M_z} \, = \, \omega_1\,M_y -\omega_1^2\tau_c\,M_z + 3\omega_1\omega_d\tau_c\,M_{zx} \nonumber\\
     & \dot{M_x} \, = \,  -\frac{9}{4}\omega_d^2\tau_c\,M_x -6\omega_1\omega_d\tau_c\,M_{yy} -3\omega_d\,M_{zy} \nonumber\\
     & \hspace{6cm} + 6\omega_1\omega_d\tau_c\,M_{zz}\nonumber\\
     & \dot{M_y} \, = \,  3\omega_1\omega_d\tau_c\,M_{xy} - (\omega_1^2 + \frac{9}{4}\omega_d^2)\tau_c\,M_y \nonumber\\
     & \hspace{5cm} - \omega_1\,M_z + 3\omega_d\,M_{zx}\nonumber\\
     & \dot{M}_{zz} \, = \, \frac{3}{4}\omega_1\omega_d\tau_c\,M_x + 2\omega_1^2\tau_c\,M_{yy} + \omega_1\, M_{zy} - 2\omega_1^2\tau_c\,M_{zz}\nonumber\\
     & \dot{M}_{xx} \, = \, 0\nonumber\\
     & \dot{M}_{yy} \, = \, -\frac{3}{4}\omega_1\omega_d\tau_c\,M_x-2\omega_1^2\tau_c\, M_{yy} - \omega_1\,M_{zy} \nonumber\\
     & \hspace{6cm} + 2\omega_1^2\tau_c\,M_{zz}\nonumber\\
     & \dot{M}_{zx} \, = \, \omega_1\,M_{xy} - \frac{3}{4}\omega_d\,M_y + \frac{3}{4}\omega_1\omega_d\tau_c\,M_z \nonumber\\
     & \hspace{5cm} - (\omega_1^2 + \frac{9}{4}\omega_d^2)\tau_c\,M_{zx}\nonumber\\
     & \dot{M}_{zy} \, = \, \frac{3}{4}\omega_d\,M_x + 2\omega_1\,M_{yy} - (4\omega_1^2 + \frac{9}{4}\omega_d^2)\tau_c\,M_{zy} \nonumber\\
     & \hspace{6cm} -2\omega_1\,M_{zz}\nonumber\\
     & \dot{M}_{xy} \, = \, -\omega_1^2\tau_c\,M_{xy} + \frac{3}{4}\omega_1\omega_d\tau_c\,M_y - \omega_1\,M_{zx}\,.
\end{align}

\subsection{Results}

We are interested in studying the dynamics of collective spin coherences and hence we choose a simple
initial condition of the form $M_x(0) = M_{\circ} \neq 0$ while the initial values of the other eight
variables in (\ref{Var}) are assumed to be zero. We analyze the dynamics of such an initial collective
coherence in the presence and in the absence of an in-phase drive of the form (\ref{dd-lockham}). The phase of
the drive is chosen to minimize first-order Rabi oscillations of the coherence and in conformity with the
experiment described by Beatrez and others \cite{beatrez21}. Using (\ref{dd-9obs}), we
then find that the two-spin dynamics is described by the following three coupled differential equations:
\begin{align}\label{dd-3d}
              & \dot{M_{x}}  =  -\frac{9}{4}\omega_d^2\tau_c\,M_{x} + 6\omega_1\omega_d\tau_c\,M^{yy}_{\rm zz} -3\omega_d\,M_{zy} \nonumber\\
              & \dot{M}^{yy}_{\rm zz}  =  \frac{3}{2}\omega_1\omega_d\tau_c\,M_{x} - 4\omega_1^2\tau_c\,M^{yy}_{\rm zz} + 2\omega_1\,M_{zy}\nonumber\\
              & \dot{M}_{zy}  =  \frac{3}{4}\omega_d\,M_{x} - 2\omega_1\,M^{yy}_{\rm zz} - (4\omega_1^2 + \frac{9}{4}\omega_d^2)\tau_c\,M_{zy},
\end{align}
where we have defined $M^{yy}_{zz} = M_{zz} - M_{yy}$. We note that the terms proportional $\omega_d^2$ in
Eq. (\ref{dd-3d}), arising from the second-order contributions of the dipolar Hamiltonian in Eq. (\ref{dd-2}),
induces damping effects in the dynamics. On the other hand, the terms proportional to $\omega_1\omega_d$,
resulting from the cross-correlations {of drive} and dipolar Hamiltonians in Eq. (\ref{dd-2}), couple the
dynamics of different two-spin variables. The initial $x$-magnetic moment $M_x(t)$ grows into the
two-spin term $M^{\rm yy}_{\rm zz}(t)$, through the drive-dipole cross-correlations. At the same time, the
drive-dipole cross-correlations convert this two-spin term into $M_x(t)$ and as such, partially compensates
for the decay of the latter. This, cycle continues until a dynamical steady-state is reached, where
$M_x(t)$, and $M^{\rm yy}_{\rm zz}(t)$ have non-vanishing values. Solving Eq. (\ref{dd-3d}), we
get the general time-dependent behavior of the collective coherence $M_x(t)$ as
\begin{equation}\label{Sol}
M_{x}(t) = M_{\circ}\left[\Big(\frac{2\omega_1}{k}\Big)^2 + 
e^{- t k ^2 \tau_c}\Big(\frac{3\omega_d}{2k}\Big)^2\cos\left(k
t\right)\right],
\end{equation}
where, $k^2 = 4\omega_1^2 + 9\omega_d^2/4$. In the presence of the drive, at large t, i.e. $t \rightarrow \infty$, 
we have a non-zero steady-state collective coherence given by 
\begin{equation}\label{SS}
       M_{x}\vert_{\rm steady\ state} =  M_{\circ}\frac{16\omega_1^2}{16\omega_1^2 + 9\omega_d^2},
\end{equation}
which we identify as the prethermal state. We have not included the system-environment coupling terms in the
analysis. These terms will lead to an unconstrained thermalization and the prethermal state will evolve to
the final nonequilibrium steady state.

On the other hand, in the absence of the drive i.e., for $\omega_1 = 0$ kilo-rad/s, the collective coherence
$M_x$, exponentially decays to $M_{x}\vert_{\rm steady\ state} = 0$. Thus it is clear that in the presence of an
in-phase external drive, we have a persistent steady-state collective coherence, which cannot be obtained
without the drive. {The magnitude of $M_x(t)$ is locked into the steady-state value after the transient
phase}, as long as the drive is kept on, {indicating the emergence of a prethermal plateau as
in \cite{beatrez21}}. Of course, in an actual experiment, this persistent collective coherence experiences a
slow decay due to the presence of $\mathcal{L}(H _{\rm SL})$ in (\ref{dd-2}). Also, out-of-phase components
of a generic drive may lead to additional decay of the signal through couplings (leakage) to the dynamics of
the other $5$ two-spin variables, which presently do not appear in (\ref{dd-3d}). {This
eventual decay of the prethermal, persistent coherence is akin to the unconstrained thermalization phase of
the dynamics of periodically driven, closed quantum many-body systems \cite{beatrez21, fleckenstein21}.}

To illustrate the behaviors, we plot the solutions of these equations using $M_{\circ} = 1$, for different
values of the drive strength $\omega_1$. We choose $\omega_d = 2\pi \times 5$ kilo-rad/s, $\tau_c = 2.5
\times 10^{-6}$ s and plot the time-series of the relevant two-spin expectation values for a period of $0$
to $50$\,ms. The chosen value of $\tau_c$ is common in (NV) centers of diamond \cite{fahobar18}. The
numerical solutions of the variables in equations (\ref{dd-3d}) are shown in figures \ref{SL0kHz} and
\ref{SL2kHz} below, both in the presence and in the absence of the drive.
\begin{figure}[h!]
\includegraphics[width=0.8\linewidth]{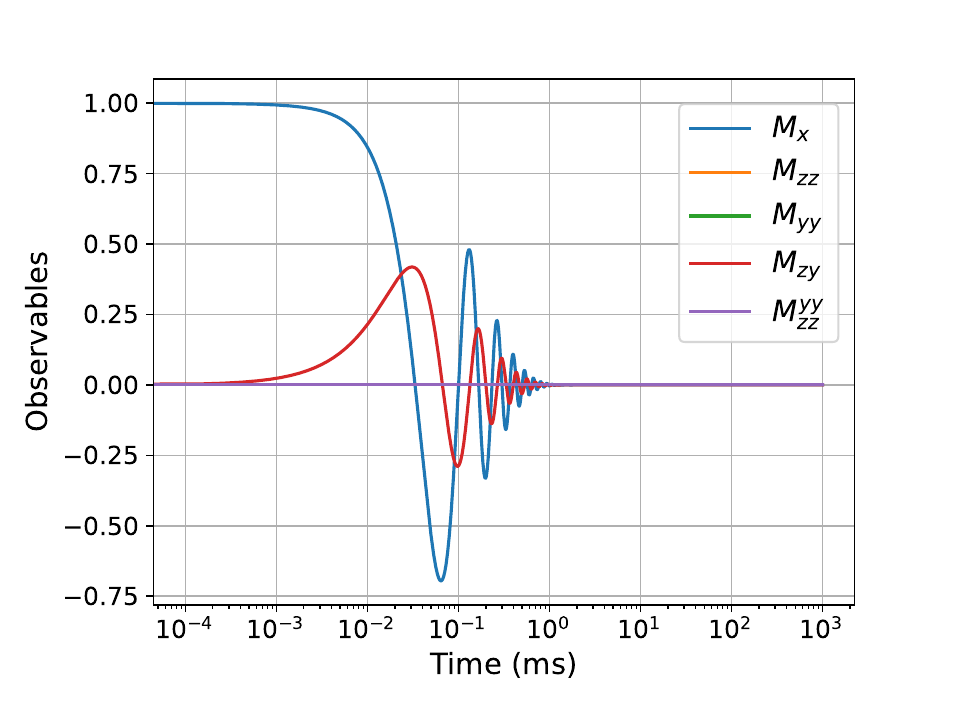} 
\caption{Dynamics of the two-spin observables in the absence of the drive i.e. $\omega_1 = 0$ kilo-rad/s. 
The time-axis is in log scale. The legends denote $M_x(t)$, $M^{\rm yy}_{\rm zz}(t)$, $M_{zz}(t)$, 
$M_{yy}(t)$ and $M_{zy}(t)$. $\omega_d = 2\pi \times 5$ kilo-rad/s.}
\label{SL0kHz}
\end{figure}
Fig. \ref{SL0kHz} shows the dynamics of the relevant two-spin variables in the absence of the drive i.e.
$\omega_1 = 0$ kilo-rad/s. In this case, after an initial {transience,}  the magnitude of $M_x(t)$ becomes
vanishingly small in the steady-state, as discussed before. Importantly, values of $M_{zz}(t)$ and
$M_{yy}(t)$ remain zero throughout the dynamics as these terms are not created from an initial collective
coherence, in the absence of the drive.

\begin{figure}[htb]
\includegraphics[width=0.8\linewidth]{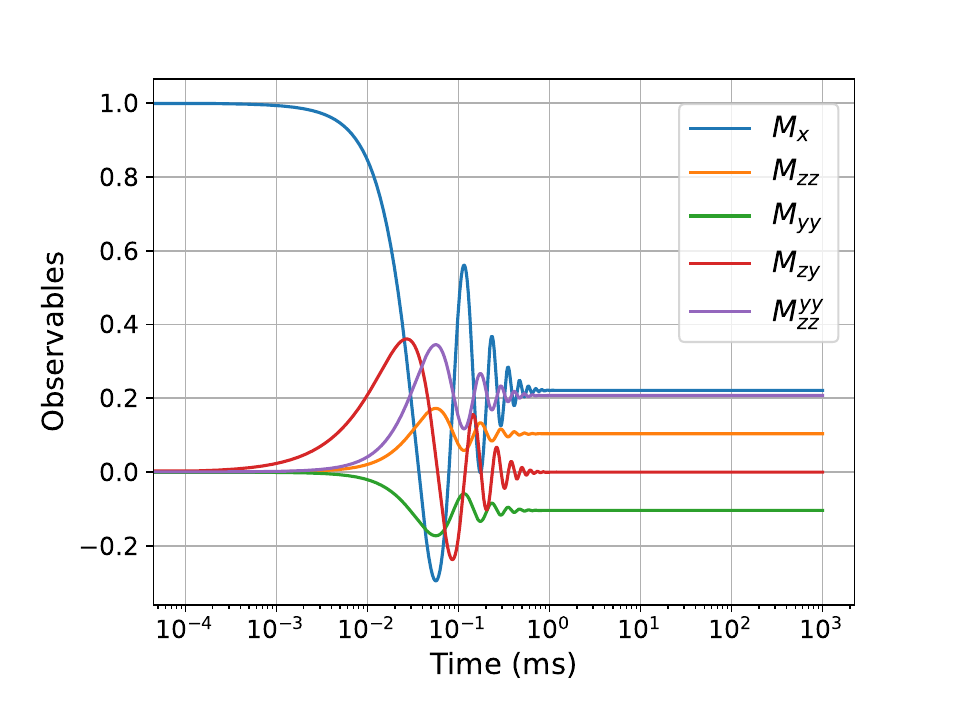}
\caption{Dynamics of the two-spin observables in the presence of a drive of amplitude $\omega_1 =
2\pi \times 2$ kilo-rad/s. The time-axis is in log scale. The legends 
denote $M_x(t)$, $M^{\rm yy}_{\rm zz}(t)$, $M_{zz}(t)$, $M_{yy}(t)$ and $M_{zy}(t)$. 
$\omega_d = 2\pi \times 5$ kilo-rad/s. Initial oscillations indicate rapid
inter-conversions between the measured values of the different observables.
The non-vanishing steady-state of $M_x$ illustrates a persistent collective coherence.}
\label{SL2kHz}
\end{figure}

The case in which the drive has a non-zero amplitude of $\omega_1 = 2\pi \times 2$ kilo-rad/s, is shown in
Fig. \ref{SL2kHz}. We find that unlike Fig. \ref{SL0kHz}, here we have a non-zero steady-state value of
$M_x(t)$ after the initial transience, illustrating the emergence of a persistent collective coherence. The
steady-state values of $M_{zz}(t)$ and $M_{yy}(t)$ are also non-zero in this case, as expected. A careful
inspection of Fig. \ref{SL2kHz} reveals that the initial $M_{x}$ gets rapidly converted to $M_{yy}$, $M_{zz}$ and
$M_{zy}$ in the transient phase. When the variables $M_{yy}$ and $M_{zz}$ have appreciable magnitude, they
get re-converted to $M_{x}$ to a large extent, resulting in the oscillatory dynamics illustrated in Fig.
\ref{SL2kHz}. Finally, the oscillations die down to result in the steady-state collective coherence. 

\begin{figure}[ht]
\centering
\includegraphics[width=0.8\linewidth]{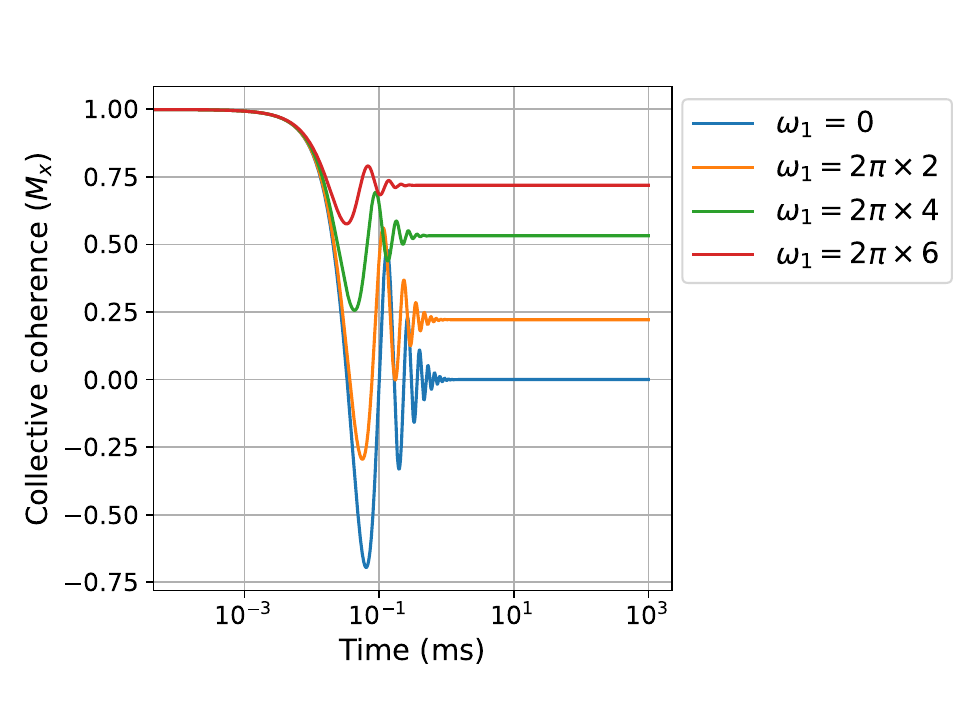} 
\caption{Emergence of the steady-state collective coherence ($M_x$) at different values of the drive strength, $\omega_1$ in kilo-rad/s. The time-axis is in log scale and $\omega_d = 2\pi \times 5$ kilo-rad/s.}
\label{SL4kHz}
\end{figure}

{To study the behavior of the two-spin dynamics for different drive amplitudes,} we plot $M_x(t)$ from the
solutions of Eq. (\ref{dd-3d}) for different values of $\omega_1$ in Fig. \ref{SL4kHz}. From Fig.
\ref{SL4kHz} it is evident that our equations predict a faster emergence of the quasi-equilibrium state with
increasing drive amplitude. Also, the transient oscillations become more prominent with decreasing strength
of the drive, as observed in the experiments of Mansfield and Ware \cite{manwar68}.  In our problem, the
damping rate of transients is obtained from equation (\ref{Sol}) as $\frac{1}{4}k^2\tau_c =
4\omega_1^2\tau_c + \frac{9}{4}\omega_d^2\tau_c$. Thus higher drive amplitudes induce faster damping of
transients (drive-induced-damping), a feature unique to the FRQME approach \cite{chakrabarti2018b}.

\section{Discussions}
Remarkably, the steady-state value of $M_x(t)$, given in equation (\ref{SS}), exactly matches the form
of the quasi-equilibrium $x$-magnetization, obtained in spin-locking experiments performed on dipolar
spin-networks.  \cite{abragam06, manwar68}. In our case, $\frac{9}{4}\omega_d^2$ plays the role of the
squared amplitude of the local field, which appears in the denominator of this quasi-equilibrium expression
\cite{abragam06, manwar68}. We note that no other QME can predict this form of the steady-state
magnetization, even though experimental confirmation of this form was obtained in the early days of magnetic
resonance spectroscopy \cite{manwar68}. It is also important to note that Mansfield and Ware's fourth-order
the perturbative approach is also incapable of predicting this result \cite{manwar68}. 

We identify the persistent coherence to be the prethermal state as reported in Beatrez and others' work
\cite{beatrez21}. If we include the system-environment coupling, we will have $T_2$ process which would
eventually lead the coherence to zero value, as an unrestrained thermalization process. We note that in the
dynamics described by Eq. (\ref{dd-3d}) there exists four conserved quantities, which are $M_{yy}+M_{zz}$,
$2\omega_1 M_x + 3\omega_d(M_{yy} - M_{zz})$, $M_{xx}$ and $\text{Tr}\{\rho_s\}$. The second of the
preceding list ensures the existence of the persistent coherence as long as $\left(2\omega_1 M_x +
3\omega_d(M_{yy} - M_{zz})\right)\vert_{t=0} \neq 0.$

The dipolar Hamiltonian (\ref{dd-sec}) transforms as a rank 2 spherical tensor while the drive Hamiltonian
is a rank 1 tensor. Cross-relaxations induced by these two Hamiltonians open up the possibility of studying
their interplay in the dynamics. Only the FRQME approach can account for these cross-correlations between
drive and dipolar Hamiltonians, which lead to the emergence of a prethermal persistent collective
coherence. Most importantly, the dynamics predicted by FRQME match with previous experimental
observations, which were not addressed by other QME techniques. 
The unique features of these cross terms (proportional to $\omega_1\omega_d\tau_c$) is that they couple two-spin 
observables of rank 1, $\sigma_x^{(1)} + \sigma_x^{(2)}$ to observables of rank 2,
$\sigma_y^{(1)}\sigma_y^{(2)}$ and $\sigma_z^{(1)}\sigma_z^{(2)}$. We note that these terms modify the rate
with which the transients decay and are responsible for giving rise to decay channels for which the the
conserved quantities emerge.

\section{Conclusions} Formulation of the FRQME for a driven, dipolar two-spin ensemble, {leads} to
cross-correlated relaxation between {drive and dipolar Hamiltonians}, in the second order. We have shown
that these cross terms {are responsible for} the emergence of a {prethermal,} persistent collective
coherence, in suitable limits. Due to the explicit presence of an exponential regulator in all second-order
terms of FRQME, arising from the average effect of fluctuations in the local environments, the
cross-correlated relaxation terms become independent of the coarse-graining interval. Other time-non-local
QME formulations, which do not have an explicit exponential regulator in the second order, can not account
for such cross-correlated relaxation effects. Thus, the bath fluctuations play a subtle but very crucial
role in the emergence of {the prethermal regime}.


\textcolor{black}{Particularly, the} agreement of our results with previous theoretical and experimental
findings indicates that the phenomenon of spin-locking in magnetic resonance can indeed be attributed to the
interplay of drive and the dipolar interactions in the second order. Our method also accounts for the
drive-dependent damping of transient oscillations observed in the {transient} phase, which could not be
explained by previous approaches. Unlike a typical magnetic resonance experiment performed on a dipolar spin
network, our present analysis is only concerned with a dipolar two-spin ensemble. However, the similarity of
our results with magnetic resonance experiments indicates the possibility of extending the present analysis
to spin-networks via a mean-field approach. On the other hand, the prethermal state of the two-spin ensemble
may be used as a short-term storage of quantum correlations. It is simple to implement, requiring a dipolar
two-spin ensemble {(which can be easily engineered)} and a resonant drive. The initial coherence can be
created through a simple $\frac{\pi}{2}$ pulse, and the drive should be in phase with this coherence. Also,
we envisage that the novel cross-terms in the FRQME may provide deeper insights into the mechanisms of
dynamic nuclear polarization (DNP) techniques, which are of considerable theoretical and practical interest
\cite{folblumahumayac09, pedfre72, maldebbajhujoomaksirvanhertemgri08, puchhoseleskta13, thawitkaucor17}.

\section{Acknowledgments} Authors gratefully acknowledge insightful discussions and critical comments on the
manuscript by Saptarshi Saha and Yeshma Ibrahim.

\end{document}